# A direct communication proposal to test the Zoo Hypothesis

João Pedro de Magalhães[1]

[1] Institute of Integrative Biology, University of Liverpool, Liverpool, UK

**Contact information:** Institute of Integrative Biology, Biosciences Building, Room 245, University of Liverpool, Crown Street, Liverpool L69 7ZB, UK; E-mail: contact@active-seti.info

**Running title:** A proactive test of the Zoo Hypothesis

# A direct communication proposal to test the Zoo Hypothesis

**Abstract:**

Whether we are alone in the universe is one of the greatest mysteries facing humankind. Given the >100 billion stars in our galaxy, many have argued that it is statistically unlikely that life, including intelligent life, has not emerged anywhere else. The lack of any sign of extraterrestrial intelligence, even though on a cosmic timescale extraterrestrial civilizations would have enough time to cross the galaxy, is known as Fermi's Paradox. One possible explanation for Fermi's Paradox is the Zoo Hypothesis which states that one or more extraterrestrial civilizations know of our existence and can reach us, but have chosen not to disturb us or even make their existence known to us. I propose here a proactive test of the Zoo Hypothesis. Specifically, I propose to send a message using television and radio channels to any extraterrestrial civilization(s) that might be listening and inviting them to respond. Even though I accept this is unlikely to be successful in the sense of resulting in a response from extraterrestrial intelligences, the possibility that extraterrestrial civilizations are monitoring us cannot be dismissed and my proposal is consistent with current scientific knowledge. Besides, issuing an invitation is technically feasible, cheap and safe, and few would deny the profound importance of establishing contact with one or more extraterrestrial intelligences. A website has been set up (http://active-seti.info) to encourage discussion of this proposal and for drafting the invitation message.

## 1. Introduction

Are we alone in the universe? Or are there other intelligent species in our galaxy? This is one of the greatest mysteries facing humankind. Given the >100 billion stars in our galaxy, many have argued that it is statistically unlikely that life, including intelligent life, has not emerged anywhere else [1, 2]. This premise led to the search for extraterrestrial intelligence or SETI, which is now over 50 years old, and was spearheaded by the Cocconi & Morrison (1959) paper and the early observations using radio telescopes by pioneers like Frank Drake [3, 4]. In spite of its thus far negative results, our search for extraterrestrial signals has barely just begun and recent advances make this effort ever more timely. Extrasolar planets are now being discovered at a rapid pace and the capacity and sensitivity of instruments for surveying the skies has been improving dramatically. For example, the Allen Telescope Array and the planned Square Kilometre Array promise unprecedented performance for SETI and for astronomical observations [5-7]. Moreover, for practical reasons, SETI has not traditionally focused on frequencies in which our civilization is more luminous but rather on primarily detecting deliberate "beacons" by other civilizations that presumably wish to signal their existence. The new generation of radio observatories will allow frequencies used for telecommunications on Earth to be surveyed as part of SETI in much greater detail [8], even if the effectiveness of such searches depends on many unknowns such as how long civilizations are "radio loud" [7]. Therefore, although there is still debate regarding the best search strategies, and funding for SETI is limited, SETI remains one of the greatest scientific enterprises of our time [3, 4].

### 1.1. Fermi's Paradox and SETI

The Milky Way is >13 billion years old and our Solar System less than half as old, suggesting that any extraterrestrial civilizations in older star systems are widely assumed to be much older and more advanced than ours [2, 3, 9-11]. The estimated time for an intelligent civilization to colonize, or at least explore, the 100,000 light year diameter galaxy is <100 million years [9]. This could involve probes, including self-replicating von Neumann probes, though there is some debate (depending on exploration strategies) regarding how long it would take to explore the galaxy [12, 13]. Be that as it may, one would expect older intelligent species to have reached us by now, and others have for long discussed the idea that extraterrestrial probes may already be in our solar system monitoring human civilization [1]. The lack of any sign of extraterrestrial intelligences, even though on a cosmic timescale extraterrestrial civilizations would have enough time to cross the galaxy, is known as Fermi's Paradox [14].

Many hypotheses have been put forward to explain this mysterious "Great Silence", including various barriers to the formation and survival of civilizations and of life itself [15]. Perhaps very few systems harbor planets suitable for life or interstellar space travel is very challenging even for advanced civilizations. While any of these explanations might turn out to be true, given our current knowledge, it is reasonable to assume that intelligent life can exist on other star systems, and interstellar travel does not violate the laws of physics and can be assumed to be practicable [16]. One additional important consideration is that the Earth has distinguishing biosignatures of life (e.g., atmospheric oxygen, water and methane in extreme thermodynamic disequilibrium) that are detectable across large distances [17]. As such, even if star travel is expensive and dangerous, and even if there are many systems to explore, the Earth has had a unique biosignature for >2 billion years [18]. Assuming that life is rare in the universe, the Earth must be a prime target for study by extraterrestrial civilizations. Therefore, an extraterrestrial civilization in our galaxy, even if modestly more advanced than ours, would likely be aware of life on our planet long enough to have reached us by now. Fermi's Paradox has thus profound implications for SETI, with historically some authors even arguing that we should abandon it, though given how little we know about the universe this appears premature [14].

## 1.2. The Zoo Hypothesis and Active SETI

Since there is no way to reliably predict the capabilities and motivations of alien civilizations, it cannot be excluded that they exist yet do not behave the way we would. One possible explanation for Fermi's Paradox is the Zoo Hypothesis, first proposed by John Ball (1973). The Zoo Hypothesis states that one or more extraterrestrial civilizations know of our existence and can reach us, but have decided not to disturb us or even make their existence known to us [19]. Many authors have debated the Zoo Hypothesis and its variants, such as the related Interdict Hypothesis [9]. The rationale behind these hypotheses is that extraterrestrial civilizations, perhaps in agreement as part of a "Galactic Club", will only contact us when we reach one or more technological, intellectual or social milestones. The possibility that extraterrestrial civilizations are lurking within the solar system or its neighborhood, perhaps observing us from the asteroid belt or from the Kuiper Belt, has been equally discussed by numerous experts. For example, it has been suggested that extraterrestrial intelligences may be observing us while deciding whether to help us or destroy us [20] or that maybe they are ignoring us without concern as to whether we detect them or not [21]. More recently, simulations have been performed addressing the Zoo Hypothesis, and in particular whether hegemony can be established in the galaxy to enforce our isolation, since all it takes is for a single discordant extraterrestrial civilization to establish contact. Results have been inconsistent, however: For example, Hair (2011) has argued that the first successful civilization in the galaxy could influence all subsequent civilization to establish a dominant cultural hegemony [22], yet Forgan (2011) has questioned these results [23],

In the context of the Zoo Hypothesis and its variants, since the 1970's many have argued that extraterrestrial intelligences monitoring us might wait for us to initiate contact and thus that we should attempt to communicate with them, but no practical way of doing this has been put forward [11]. Active SETI, also called messaging to extraterrestrial intelligence or METI, is the attempt to send messages to extraterrestrial civilizations. It has been extremely controversial within the SETI community since the first historical *Arecibo Message* was sent in 1974 aimed at a distant star cluster [10]. The major concern is that sending interstellar messages could reveal our location to potentially hostile extraterrestrial civilizations. This has not stopped various Active SETI attempts, though, including the more recent *Cosmic Calls* messages and the *Teen Age Message* [reviewed in 10]. One notable attempt related to the Zoo Hypothesis was made in the form of the Invitation to ETI website (http://ieti.org/) led by the late Allen Tough. The idea behind this website was for it to act as an invitation to extraterrestrial civilizations already observing humankind to contact us (e.g., by e-mail). Its assumption, however, is that such alien civilizations monitoring us can access and interact with our Internet, which is highly dubious (or unproven at best) since this would require a connection (i.e., transmitting and receiving data) with a computer on Earth. Therefore, there is an unmet need to develop an Active SETI protocol in the context of the Zoo Hypothesis.

## 2. An Active SETI proposal to test the Zoo Hypothesis

I propose here a proactive test of the Zoo Hypothesis. Specifically, I propose to send a message to any extraterrestrial intelligence(s) that might already be observing us and inviting them to respond. My aim is to attempt to bring forward the communication with extraterrestrial civilizations by stating that we are ready to engage with them at a high level. The rationale is that, assuming the Zoo Hypothesis is true, extraterrestrial civilizations must be observing human civilization, which must involve monitoring our radio leakage as this is readily detectable at long distances (i.e., from outside the solar system) [24]. My assumption is also that it is possible to influence the decision-making process of extraterrestrial civilization(s), by initiating contact or perhaps (as detailed below in section 2.3, the exact content of the invitation message is still to be determined) by asking for their assistance. Because we frequently regard past human societies (even from a recent past) as primitive, it is certainly possible that a more advanced species would still consider present human values and social structure as unfit for any sort of communication. After all, future human generations are likely to regard our thinking now as incorrect and even backwards. It is also possible that extraterrestrial intelligences consider certain aspects of our biology as unsuitable for engaging with them, like our short lifespans that could prevent them from communicating with the same individual if communications take a long time by human standards. That said, and in spite of the unlikeliness of the many assumptions underlying my proposal (starting with the Zoo Hypothesis itself), I believe this is a worthwhile endeavor since it can be achieved with very modest resources (see section 2.2 below). Besides, the prospect of being successful, no matter how unlikely this is, is tantalizing since establishing contact with extraterrestrial intelligences would forever change humanity. The act of sending such a message will, by itself, energize SETI and force us to more profoundly consider the prospect of communicating with extraterrestrial civilizations.

### 2.1. Potential drawbacks of establishing contact with extraterrestrial intelligences

Opposition to direct communication efforts is mostly based on concerns related to costs and the potential dangers of revealing ourselves [10, 11, 25, 26]. Many authors have pointed out that we have good reasons to believe that extraterrestrial intelligences would be capable of destroying our civilization [25], or at least pose considerable risks [27]. (Neal (2014) argues that we should minimize such risks by improving international collaboration and even military readiness in view of the prospects of contacting extraterrestrial intelligences.) If extraterrestrial civilizations are already aware of us and eavesdropping on us, however, then attempting to communicate with them will not put us in any danger, at least not in any more danger than we are already. To put it another way, if advanced extraterrestrial intelligences can reach our solar system it is reasonable that they have likely chosen not to destroy us yet, which would indicate that they are more prone to be cooperative. Although predicting the behavior, motivation and agenda of extraterrestrial civilizations is impossible, others have argued that cooperative extraterrestrial intelligences are more likely to be extant in the galaxy [25], though the possibility that they are neutral or even malevolent cannot be excluded [10]. Even if extraterrestrial civilizations are cooperative and aim to help us, some disruption of our society (e.g., cultural shock) is to be expected, even if we assume that the benefits of engaging extraterrestrial intelligences outweighs its dangers. Likewise, unintended negative outcomes from engaging extraterrestrial intelligences, such as the spread of new diseases or military applications of extraterrestrial technology [10], is a possibility, even though I feel these are unlikely if indeed extraterrestrial intelligences are cooperative and intent on helping us. Importantly, if advanced extraterrestrial civilizations intend to help us at some point then communicating with them sooner rather than later will benefit our species. Since, as detailed below (see section 2.2), my proposal involves standard radio and television that are already being transmitted, it will not expose our existence anymore than we are doing already. Therefore, and unlike traditional direct communication efforts aimed at other systems, which have been a source of controversy [25], there is little risk in my proposed endeavor.

**2.2. Transmitting the invitation via radio and television broadcasting**

I propose to use radio and television broadcasting for transmitting the invitation message. Sullivan (1978) surveyed the radio signature of the Earth and found that this is detectable at interstellar distances. In fact, television and radio broadcasts are (apart military radars) the most detectable of our radio leakage [24]. More recently, Sagan et al. (1993) found empirical evidence using data from the Galileo spacecraft that television and radio transmissions can be detected off orbit. Similarly, the Wind spacecraft on orbit detected radio transmissions from the Earth [28]. Therefore, and in the context of the Zoo Hypothesis, an extraterrestrial intelligence could be eavesdropping on us from a distance, even from other star systems (but see below). While the limits of detection of Earth's radio transmissions are a subject of debate (Sullivan argues ~25 light-years, Atri et al. (2011) and Baum et al. (2011) up to 100 light years), as they largely depend on the size of the receiving antenna, the crucial point is that an extraterrestrial intelligence would be able to gather a wealth of information simply by eavesdropping on our radio and television broadcasting. This might even be achieved without the need for physical presence in the solar system, although this is debatable since the signal-to-noise ratio decreases with distance and, for instance, Kaiser et al. (1996) have argued that detecting man-made signals is not possible from other star systems unless with extremely large antennas. An additional concern is that as the signal-to-noise ratio decreases with increasing distance from the Earth, our radio leakage may no longer be decipherable, even if it remains detectable. Be that as it may, I propose to use existing radio and television broadcasting to send our invitation to any extraterrestrial civilization(s) that might be listening. When compared to the Invitation to ETI website, this also has the significant advantage that there is empirical evidence that our television and radio leakage can be detected off orbit and, besides, these signals have been going on for decades.

      Given that with our current technology we have the ability to analyze and interpret multiple and complex radio transmissions, it is reasonable to assume that an advanced extraterrestrial civilization eavesdropping on us could easily detect and interpret our radio and television leakage. Sullivan (1978) does point out some physical limitations. In particular, AM broadcasting, in contrast to FM, does not normally escape the ionosphere. Therefore, this affects our choice of transmission method. We should also choose a transmission that has been going for a long time, ideally in North America or Western Europe peaks as these have been for decades producing repeatable signals in our 24-hr cycle [24]. In addition, UHF is more likely to be detectable [24]. Consideration should also be given to season, diurnal variations and locations with low ionospheric plasma density as this prevents radio waves from escaping into space [28]; transmissions in the winter, just before sunrise and at higher latitudes should be favored.

      With the above considerations in mind, most radio and television stations would be suitable; of course, broadcasting stations that use transmissions via satellite or cable will be

excluded, and ideally the same exact message would be sent via various stations (and keeping in mind that each station broadcasts simultaneously from several transmitters) and in both video and audio. Given the large number of stations transmitting worldwide, often using the same frequencies [28], transmitting from multiple stations and therefore in multiple frequencies would increase the probability that our transmission will stand out. Similarly, although I anticipate that the message will be in English, it will ideally also be broadcast in other languages. Since we will be using standard radio and television transmissions, signals will be transmitted in a fairly isotropic way [8]. Modulation and encoding should be (like human language) understandable to more advanced extraterrestrial civilizations that have been observing us presumably for quite some time. (One of the advantages of transmitting in the context of the Zoo Hypothesis is that it decreases cultural and communication barriers.) Transmitter specifications will vary widely amongst the thousands around the world, but radio and television frequencies within 40-850 MHz are anticipated [8]. These frequencies already exclude AM broadcasting and preference should be given to top-of-the-range UHF frequencies. The power is typically above 1 kW per television or radio transmitter and maximum power is possibly over 500 kW [8], but this is highly variable across radio and television stations and even across the multiple transmitters of each station. An ideal broadcast would consist of a combination of older transmitting stations, powerful transmitters and as many different frequencies as possible. In addition to using multiple stations, we should also regularly repeat the transmission, as pointed out by others [29]. Lastly, it is vital that we keep appropriate records of all transmissions (i.e., broadcasting stations with frequencies, locations, time of transmission, etc.) for future reference (see section 2.6 below on long-term prospects).

**2.3. Drafting an invitation message**

Perhaps, as argued by others [reviewed in 25], simply issuing an invitation is the necessary milestone for extraterrestrial intelligences to respond. On the other hand, we would probably do well to try to convince extraterrestrial intelligences to engage with the human species and make it as easy as possible for them to do so. One major open question, thus, concerns the content of our message. I believe this invitation message should: 1) acknowledge the Zoo Hypothesis and express our wish to communicate with extraterrestrial intelligences as soon as possible; 2) with limited resources on Earth, a growing population and the capacity for self-destruction (e.g., due to nuclear weapons), our civilization is in more danger than ever; therefore, we may well need help to survive and this is why we would like to learn from older, more advanced civilizations; and 3) suggest an easy way for extraterrestrial civilizations to respond to us (see section 2.4 below). A website concerning this project has been set up (http://active-seti.info) to organize this effort and to keep a record of transmissions. Suggestions for how best to tailor our message and expressions of interest in this project are welcomed. My aim here is to start a rational, scientific discussion on the content of the message to be transmitted in due course. I anticipate that a group of interested parties (possibly mostly scientists, but nonscientists are welcomed) will draft and, following public consultation on the project website, approve the final message.

Because transmitting an invitation with my proposed method is so simple and cheap, I wish to establish a common front on this topic, otherwise many groups could start issuing invitations which may be counterproductive. The point has been made, however, that universal consensus is either impossible or will result in a poor representation of the diversity of humankind [26]. Besides, several topics that can be covered in the invitation message are controversial in that they may increase the likelihood of engagement from extraterrestrial civilizations while at the same time reducing the broader appeal of our message to humankind, for example in terms of wishing to engage extraterrestrial intelligences on a purely scientific basis that rejects religion. My view is that some cohesion is necessary to ensure that humankind is adequately represented but also that the invitation message is fit for purpose in terms of increasing our chances of enticing a response from extraterrestrial intelligences. Hopefully, a suitable discussion will follow from this paper to clarify these issues. Likewise, other elements could be included in the message and suggestions are welcomed. For example, it has been argued that extraterrestrial intelligences are likely to value life in general or perhaps complex animals, but not humans in particular or any of our anthropocentric values [25]; if so, we could express a willingness to respect other complex life forms, even if this is at odds with many of our current activities, and as suggested by others [14] extraterrestrial intelligences could be postbiological or machine servants. Moreover, should further details of human biology and society be included in the message, and, if so, which details? Should we try to paint a more favorable picture of human civilization? Or should we assume that any eavesdropping extraterrestrial civilization will already be deeply knowledgeable of us? These and other points discussed above are open to further debate.

**2.4. Suggested method for extraterrestrial civilizations to respond to us**

For simplicity, extraterrestrial intelligences should respond to us by using one or more of the frequencies used to transmit the invitation message. This would ensure that, since these are normal radio and television broadcasting services, they would be noticed by us. There is an issue of whether extraterrestrial entities will have the capacity to transmit a strong enough signal, possibly from outside the solar system or maybe even from nearby star systems, to interfere with our radio or television transmissions in what are clearly noisy frequencies. An alternative would be to suggest a less noisy frequency, like the 21 cm hydrogen line, though this would require us to set a way of monitoring such a frequency at a given set time(s) and that that the response from extraterrestrial intelligences be sent at the suggested time(s). Since it is impossible to predict the circumstances surrounding any extraterrestrial entities (from a small probe to large staffed spaceships, perhaps even from different factions or species, or as postbiological entities), we should also suggest that they respond to us in any way they see fit.

**2.5. Response to extraterrestrial intelligences**

Even if unlikely (I acknowledge that extraterrestrial intelligences are probably not watching us), we should be prepared for the prospect that extraterrestrial intelligences will respond to our invitation and a subsequent appropriate response to extraterrestrial intelligences must be established. Fortunately, this aspect of my proposal has already been dealt with extensively in the context of SETI in terms of establishing an appropriate reception of signal and response [30]. The established protocol is not to respond immediately and instead inform the United Nations and its Office for Outer Space Affairs for them to decide, following international consultations, on an appropriate response that reflects the broad concerns and wellbeing of humanity. Although there is still some debate concerning the exact protocol, the SETI reply protocol will also be followed in my proposal. One aspect of my proposal that makes it particularly timely is that our civilization's communications ability have an unprecedented capacity, meaning that we now have the ability, for the first time in human history, to quickly disseminate information and communicate across the planet. One potential drawback is that if the extraterrestrial response to our invitation is received as an open signal (as proposed in 2.4 above), no doubt many states and groups would have the capacity to respond and this is likely to result in a cacophony of responses, a caveat that has also been made in context of SETI [3]. Extraterrestrial intelligences could, of course, choose how to communicate with us, which makes this very difficult to predict.

**2.6. Interpreting negative results and long-term prospects**

A null result (i.e., no response received from extraterrestrial intelligences) could mean that: 1) there is no extraterrestrial civilization listening (either because we are alone in the galaxy or because they have not reached us yet); 2) extraterrestrial intelligences are watching us but did not receive our message; 3) extraterrestrial intelligences received our message but chose to ignore it. The latter is the most interesting hypothesis as it would mean that we would have failed to persuade them, perhaps because of unconvincing arguments or because other milestones must be reached by humankind. One important implication, however, is that assuming that our message includes an explicit request for assistance (as I think it should; see 2.3 above), this would change our view towards extraterrestrial intelligences in the context of the Zoo Hypothesis. Thus far the assumption is that extraterrestrial intelligences have been watching us without interfering but for the first time they would have made the decision not to respond to our request for assistance. If the Zoo Hypothesis later turns out to be correct, the failure of extraterrestrial intelligences not to assist us when requested will require some justification and may even influence future diplomatic relations.

Of course, even in the event of a null result, it is difficult to conclude that there is no extraterrestrial intelligence observing us. This is a limitation also encountered by traditional SETI efforts which have been argued may continue for hundreds or thousands of years until we can conclude that there is nobody out there [3]. We may eventually reach a point where we can disprove the Zoo Hypothesis, but for the foreseeable future an absence of a response from extraterrestrial intelligences does not imply an absence of extraterrestrial intelligences monitoring us. Therefore, one option would be to have further transmissions, perhaps recapping important advances in humankind. In this way, future researchers can decide whether humankind has progressed enough for us to transmit again and/or whether technical advances warrant new transmissions. Regular Active SETI transmissions have indeed been proposed as a way to invigorate the field [11], and this way we could have a systematic, long-term strategy for transmitting and for coordinating such efforts. In this context, an advantage of my proposal is that it is relatively cheap; no new equipment or facilities are necessary, and transmissions could be performed at times when radio and television stations are not normally broadcasting to minimize costs. Keeping costs to a minimum is essential since, even in the broader context of SETI, funding is limited. Adequate recordkeeping of transmissions may then be necessary for hundreds or thousands of years. While it is impossible to predict the future of human civilization on such large timescales, electronic records that can be easily copied and kept by different individuals at different physical locations seem to me to be most appropriate. With sufficient individuals and resources, an institution or organization can even be set up to coordinate efforts and permit long-term institutional memory.

**3. Concluding remarks**

In conclusion, I propose a proactive test of the Zoo Hypothesis consisting of sending a message using radio and television to any extraterrestrial civilization that might be eavesdropping on us with the goal of engaging with them for mutual benefit. Even though I accept this is unlikely to be successful in the sense of resulting in a response from extraterrestrial intelligences, the possibility that extraterrestrial civilizations are monitoring us cannot be dismissed and my proposal is consistent with current scientific knowledge [1, 2, 24]. Importantly, we are already broadcasting radio and television signals without editorial control from humankind, and any eavesdropping extraterrestrial civilizations detecting them will receive a biased view of our species based mostly on how it entertains itself. Broadcasting an explicit invitation message may improve the odds of extraterrestrial intelligences communicating with us and will not increase our risks. Besides being safe, issuing an invitation is technically feasible, cheap and with tremendous potential benefits since few would deny the profound importance of establishing contact with one or more extraterrestrial intelligences [2]. Communicating with more advanced extraterrestrial civilizations would forever change humankind.

**Acknowledgments:**

I wish to thank William Baines and Duncan Forgan for critical comments on previous drafts and Thomas Craig for help setting up the project website. This proposal was presented at the third annual meeting of the UK SETI Research Network in Leeds, UK, September 2015, and I wish to further thank all the participants for fruitful discussions on this topic. Work in my lab is funded by the Wellcome Trust and the UK Biotechnology and Biological Sciences Research Council (BBSRC).